\documentclass[aps, prb, reprint, superscriptaddress, floatfix, longbibliography]{revtex4-2}
\usepackage{graphicx, amsmath, amsfonts}
\usepackage{ulem, lipsum, verbatim, hyperref}
\usepackage{multirow, xcolor}

\definecolor{TODO}{HTML}{cccc00}
\definecolor{IDEA}{HTML}{33cc00}
\definecolor{DOUB}{HTML}{3333ff}

\newcommand{\vb}[1]{\mathbf{#1}}
\newcommand{\mean}[1]{\left\langle #1\right\rangle}
\newcommand{\abs}[1]{\left\lvert #1\right\rvert}

\begin{document} 

\title{Global sampling of Feynman's diagrams through Normalizing Flow}
\author{Luca Leoni}
\email{luca.leoni12@unibo.it}
\affiliation{Department of Physics and Astronomy 'Augusto Righi', Alma Mater Studiorum - Universit\`{a} di Bologna, Bologna, 40127 Italy}

\author{Cesare Franchini}
\affiliation{Department of Physics and Astronomy 'Augusto Righi', Alma Mater Studiorum - Universit\`{a} di Bologna, Bologna, 40127 Italy}
\affiliation{University of Vienna, Faculty of Physics and Center for Computational Materials Science, Kolingasse 14-16, 1090 Vienna, Austria}
\date{\today}

\begin{abstract}
	Normalizing Flows (NF) are powerful generative models with increasing applications in augmenting Monte Carlo algorithms due to their high flexibility and expressiveness.
	In this work we explore the integration of NF in Diagrammatic Monte Carlo (DMC), presenting an architecture designed to sample the intricate multidimensional space of Feynman's diagrams through dimensionality reduction.
	By decoupling the sampling of diagram order and interaction times, the flow focuses on one interaction at a time. This enables constructing a general diagram by employing the same unsupervised model iteratively, dressing a zero-order diagram with interactions determined by the previously sampled order.
	The resulting NF-augmented DMC is tested on the widely used single-site Holstein polaron model in the entire electron-phonon coupling regime.
	The obtained data show that the model accurately reproduces the diagram distribution by reducing sample correlation and observables statistical error, constituting the first example of global sampling strategy for connected Feynman's diagrams in DMC.
\end{abstract}

\maketitle

\section{Introduction}
The investigation of many-body effects in interacting systems has long presented a significant challenge in the physics community.
A modern approach to address this challenge centers on the strategic use of Feynman's diagrams for perturbative assessments of key quantities like self-energies and correlation functions~\cite{feynman1949, mahanManyParticlePhysics2000}.
These evaluations provide insights into the intricate ways in which single-particle properties are impacted by interactions. Consequently, this approach establishes a comprehensive framework with versatile applications extending to thermodynamics and quasi-particle properties, thereby attracting interest from a variety of fields~\cite{stefanucci2013}.

Diagrammatic quantum Monte Carlo (DMC) is a powerful numerical method to obtain approximation-free estimates of diagrammatic perturbative expansions~\cite{prokofev1998, mishchenko2000}.
A target quantity, usually the interacting Matsubara Green's function $G(\vb{k}, \tau)$, is sampled stochastically through the Markov chain Monte Carlo (MCMC) exploration of the Feynman's diagrams contributing to its power series.
This approach has been widely used in condensed matter physics to accurately solve a wide range of physical systems described in terms of effective Hamiltonians.
Representative examples are given by polarons~\cite{franchiniPolaronsMaterials2021, prokofev1998, mishchenko2000, Hahn2018, mishchenko2003, PhysRevB.104.L161111, PhysRevB.107.L121109}  and excitons~\cite{Burovski2001, Burovski2008}, thermodynamic of spin systems~\cite{kulagin2013, defilippis2021} and correlation of electron gas~\cite{chen2019}.
While considered to be numerically exact, DMC becomes progressively less efficient with increasing complexity of diagram's phase space, which rapidly scales when approaching real materials.
For example, the incorporation of ab-initio material-specific band dispersions and electron-phonon coupling introduces distinctive features in the phase space, leading to a sluggish MCMC exploration with increased autocorrelation between samples, inducing critical slowing down effects~\cite{wolff1990}.
Other Monte Carlo algorithms tackle this issue by employing global sampling strategies, which enhance the exploration of phase space within MCMC~\cite{wolff1989}.
These approaches, however, are not yet present in the DMC literature~\cite{prokof15}.
Nevertheless, recent advancements in generative machine learning (GML) have demonstrated the potential for such architectures to serve as effective global updates in various physics-related scenarios~\cite{albergo2019, singha2023, singha2023a, nicoli2021, huang2017, gabrie2022, cranmer2023}.

From a broader perspective, the integration of ML techniques in many-body physics has gathered substantial interest in recent years~\cite{carleo2019}.
Noteworthy successes in performance enhancement have been achieved across various contexts, encompassing density functional theory~\cite{moreno2020, snyder2012, brockherde2017}, numerical renormalization group~\cite{li2018, disante2022}, and interacting spin models \cite{carleo2017}.
Particularly relevant to the current context is the integration of GML models into Monte Carlo schemes to refine sampling from high-dimensional, intricate probability distributions~\cite{wu2019, noe2019}.
Various GML algorithms have proven effective as global updates~\cite{wolff1989, prokofev2001, wolff1988}, enhancing MCMC procedures~\cite{huang2017, gabrie2022, cranmer2023}. This enables the generation of uncorrelated samples within the chain and mitigates the critical slowing down problem.

Normalizing Flows (NF) represent an emerging family of generative models, standing out as one of the most promising techniques in GML. NF provides a method to approximate invertible maps to desired probability distributions~\cite{papamakarios2021, kobyzev2021}. The NF architecture facilitates both rapid sampling and inference of target distributions, exhibiting significant versatility in representing complex objects. It has found successful applications across a wide spectrum, including image and video generation, reinforcement learning and
in the Monte Carlo integration of quantum lattice field models~\cite{albergo2019, singha2023, singha2023a, nicoli2021}.
The prospect of designing a flow-based model to capture the features of the Feynman diagram's phase space within a DMC framework is highly appealing. However, the added complexity in this case lies in the fact that the sample's dimensionality is a variable of the distribution, rendering standard NF architectures unsuitable for addressing the problem. In this letter, we introduce an approach to circumvent such limitations by proposing a model that generates a diagram iteratively, focusing on individual interactions. To evaluate the feasibility and efficiency of this hypothesis, we employ the single-site polaron Holstein model as a test case, a paradigmatic electron-phonon (el-ph) effective Hamiltonian~\cite{holstein1959}.

Our designed NF augmented DMC approach yields a significant reduction in sample correlation within the Markov chain across a broad spectrum of electron-phonon coupling strengths. This reduction translates into a diminished number of samples required to achieve convergence. The proposed data-driven global updates enhance statistical convergence, leading to a roughly 50\% reduction in both statistical errors associated with observed polaron properties and sample autocorrelation compared to the standard DMC local procedure. Since the code developed in this study was designed to enhance the exploration of Feynman's diagram space rather than to optimize performance, the actual sampling cost is higher compared to the standard approach.

The article is organized as follows. In the next section, we introduce the NF-augmented DMC architecture tailored for the Holstein polaron model in the atomic limit. Subsequently, we present and analyze the performance of the developed code in estimating specific polaron quantities.

\section{Normalizing Flow Diagrammatic Monte Carlo}

\subsection{Diagram representation}

The single site Holstein model is expressed as
\begin{equation}
	H = -\varepsilon \hat{c}^\dagger\hat{c} + \Omega\sum_{\vb{q}} \hat{b}^\dagger_{\vb{q}}\hat{b}_{\vb{q}} + \frac{g}{\sqrt{N}} \sum_{\vb{q}} \hat{c}^\dagger\hat{c}(\hat{b}^\dagger_{-\vb{q}} + \hat{b}_{\vb{q}}),
\end{equation}
consisting of a single non-dispersive electronic band with energy $\varepsilon$, interacting with an Einstein-like phononic branch of frequency $\Omega$. The electron-phonon coupling is characterised by a quadratic term with a constant interaction vertex $g$, and $N$ represents the number of possible phonon states.

\begin{figure}[t]
	\centering
	\includegraphics[width=\linewidth]{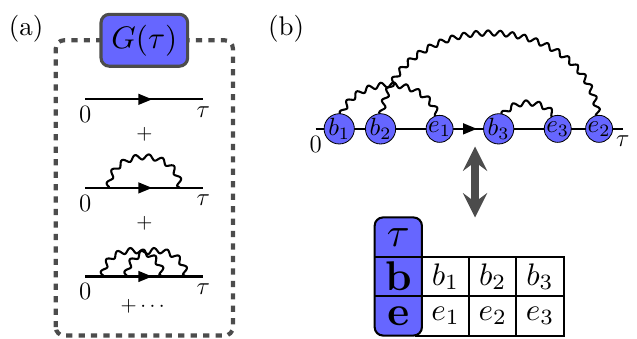}
	\caption{(a) Graphical representation of the Green's function expansion for the single site Holstein model. (b) Vector description of a general diagram used in this work, the interaction times are stored in two separate vectors collecting the beginning and the end of the phonon lines while $\tau$ encode the length of the electron one, defining the diagram's length.}
	\label{fig:DiagExp}
\end{figure}

We are interested in the quasi-particle properties of the model, which are encoded in the interacting Green's function.
For polaronic systems the Green's function can be expressed as a sum over diagrams having an increasing number of phonons, as sketched in Fig.~\ref{fig:DiagExp}(a).
Using Feynman's rules we can translate such expansion in a mathematical expression that, in the Matsubara formalism, takes the form:
\begin{equation}
	\label{eq:Green}
	G(\tau) = e^{-\epsilon\tau}\sum_{n=0}^{\infty} \int_0^{\tau}\text{d}x_1\int_{x_1}^{\tau}\text{d}x_2\cdots\int_{x_{2n-1}}^{\tau}\text{d}x_{2n} W(\vb{x}).
\end{equation}
The ordered integration accounts for every diagram having $n$ phonons, each described by a vector of $2n$ interaction times $\vb{x}$.
The contribution of each diagram is expressed by the zero-temperature electronic and phononic propagators, resulting in the following diagram's weight:
\begin{equation}
	\label{eq:Weight}
	W(\vb{x}) = g^{2n}\prod_{i=1}^n e^{-\Omega(e_i - b_i)},
\end{equation}
where $\vb{x}$ was divided in phonon's creation, $\vb{b}$, and annihilation times, $\vb{e}$, allowing the diagram to be represented by a vector $[\tau, \vb{b}, \vb{e}]$ (see Fig.~\ref{fig:DiagExp}(b)).

DMC directly integrates Eq. \ref{eq:Green} by randomly sample expansion's diagrams from a distribution  proportional to their weight function.
Detailed information on how the MCMC procedure is implemented in diagram space are found at \cite{vanhoucke2010, greitemann2018}, while the observable estimators for polaron quantities are introduced by Mishchenko~\cite{mishchenko2000}.
Here we focus our attention on $G(\tau)$ and the polaron binding energy $E_p$.
The Green's function is evaluated by the histogram collecting the distribution of the variable $\tau$, while $E_p$ through the ground state energy estimator~\cite{mishchenko2000}:
\begin{equation}
	E_p = \lim_{\tau\to\infty} \mean{\frac{\Omega\sum (e_i - b_i) - 2n}{\tau}}.
\end{equation}
Analytical expressions for both observables can be found in the literature~\cite{mahanManyParticlePhysics2000}, making them a suitable choice for testing the convergence properties of a new algorithm.

\subsection{Flow architecture}
\label{sec:FlowArch}

Normalizing flows transform a simple starting probability distribution, $p_Z$, into a more complex desired density, $p_X$, through a series of invertible and smooth transformations.
This is achieved using invertible networks~\cite{germain2015, dinh2017, kingma2018} to construct a variational map $T_{\boldsymbol{\theta}}: \mathbb{R}^D\to\mathbb{R}^D$ that, when applied to samples $\{\mathbf{z}_i\}_{i=0}^N$ drawn from $p_Z$, generates a set $\mathbf{x}_i = T_{\boldsymbol{\theta}}(\mathbf{z}_i)$ distributed as~\cite{papamakarios2021}:
\begin{equation}
	\label{eq:NF}
	p_{\boldsymbol{\theta}}(\vb{x}) = p_{Z}\circ T^{-1}_{\boldsymbol{\theta}}(\vb{x}) \left\lvert\det J_{T^{-1}_{\boldsymbol{\theta}}}(\vb{x})\right\rvert,
\end{equation}
with $J_{T^{-1}_{\boldsymbol{\theta}}}$ being the Jacobian of the invertible map.
This provides fast sampling and density estimation of a parametrized distribution $p_{\boldsymbol{\theta}}(\mathbf{x})$, which can be fitted to the target $p_X(\mathbf{x})$ by minimising their Kullback-Leibler (KL) divergence.

\begin{figure}[t!]
	\centering
	\includegraphics[width=\linewidth]{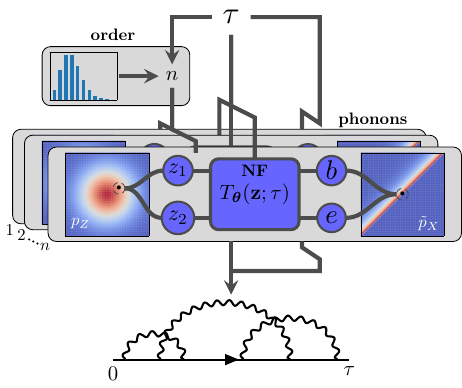}
	\caption{Representation of the NF-DMC workflow. The model takes as input the diagram's length $\tau$, which is used as conditional parameter to sample both the order $n$ and phonon interactions times \textbf{b} and \textbf{e}. $n$ is initially sampled from a suited distribution (e.g. the Poisson distribution shown in histogram form in the inset).
		Then an increasing number of phonon lines is added to the initial zero-order diagram ($n=0$) defined by $\tau$ by transforming gaussian samples \textbf{z} using the $\tau$-conditioned NF transform $T_{\boldsymbol{\theta}}$.
		Finally, for each $\tau$ an $n$-order diagram (i.e. with $n$ phonon lines) is obtained, as sketched in the bottom.
	}
	\label{fig:modelArch}
\end{figure}
For DMC the target $p_X(\mathbf{x})$ is the diagram's weight distribution proportional to the weight $W(\mathbf{x})$, with $\mathbf{x}$ being the diagram's vector representation introduced in Eq.\ref{eq:Weight} and Fig.\ref{fig:DiagExp}(b).
The complexity of approximating the full $p_X(\mathbf{x})$ in a NF framework lies in the dimensionality of the vector space spanned by the different diagrams. In fact, $\mathbf{x}$ changes dimensions based on the number of phonons, described by $n$ (see Eq.~\ref{eq:Green}), typically referred to as the diagram's order. Approaching this problem with standard NF architectures would mean creating a different model for every possible dimension $n$, leading to the impossibility of covering the entire space.
Here, we avoid such restriction by decoupling the sampling of the diagram's order and the interaction times with the approach schematised in Fig.~\ref{fig:modelArch}.

The strategy relies on sampling single phonons using a conditional NF model $T_{\boldsymbol{\theta}}$~\cite{singha2023}, with specifics reported in appendix~\ref{App:DesOrDis}, trained to fit a 2D distribution ($\tilde{p}_X$) describing $b$ and $e$ statistics based on the current diagram's length, $\tau$.
Thus, allowing the generation of a new diagram by multiple use of $T_{\boldsymbol{\theta}}$ to dress the zero order diagram, defined by $\tau$, with a random number of phonons independently sampled from a preselected integer distribution $p_n(n \vert \tau)$.
In this way, the entire diagram space at different $\tau$ can be explored using a single model possessing the following final density:
\begin{equation}
	\label{eq:ModelDensity}
	p_W(\vb{x}) = p_n(n \vert \tau)\left[n!\prod_{i=1}^n p_{\boldsymbol{\theta}}(b_i, e_i\vert \tau)\right],
\end{equation}
where the $n!$ is needed since every permutation of the phonons indices leads to the same diagram.
The combination of this density with Eq.~\ref{eq:NF} can be used to completely define an update in the DMC chain.

Finally, the model was completed by selecting $p_n(n\vert \tau)$ as a Poisson distribution exhibiting a mean value $\lambda$ dependent on both $\tau$ and the el-ph coupling strength $g$. The form of $\lambda$ was chosen to best fit the order statistic collected on preexisting runs at different $\tau$ with varying $g$, and showed in appendix~\ref{App:DesOrDis}.
This leads to a final $p_n$ conditioned also by the coupling constant $g$. For instance, sampling from different coupling regimes comes at no additional increase in the NF complexity due to the $g$ dependence being completely contained in the simpler order distribution.

\subsection{Loss function}

\begin{figure}[t!]
	\centering
	\includegraphics[width=\linewidth]{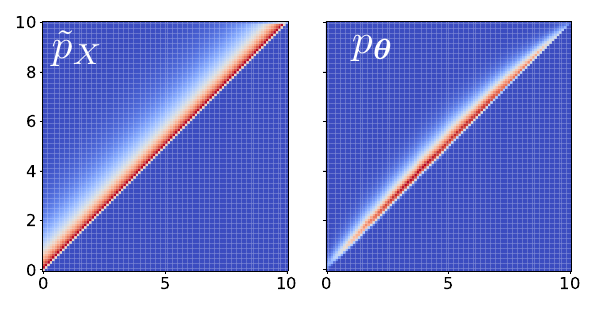}
	\caption{Target density $\tilde{p}_X$ and the obtained reconstruction using the NF model $p_{\boldsymbol{\theta}}$. Both are plotted for a selected value of the conditional paramter $\tau$ of 10, while the hyperparameter $\alpha$ was set to 50.}
	\label{fig:TrainRes}
\end{figure}

In the considered Holstein model, comparing Eq.\ref{eq:ModelDensity} with the target weight in Eq.\ref{eq:Weight} shows that the model distribution $p_{\boldsymbol{\theta}}$ should be trained to fit the unnormalized density $p_X(b, e) = e^{-(e - b)}$, where $\Omega$ is set as the unit of measure for the energies. Also, to generate a physical phonon the flow sampling needs to take into account that the annihilation $e$ must happen after the creation $b$ and both of them needs to be placed before the end of the diagram.
We found that the most flexible way that allows $e$ and $b$ to satisfy such constraint is multiply the target $p_X$ by steep sigmoid functions to create a new $\tau$-dependent target $\tilde{p}_X$ with vanishing contribution in the unphysical regions of the domain
\begin{align}
	 & \tilde{p}_X(b, e; \tau) = p_X(b,e)\sigma(b)\sigma(e-b)\sigma(\tau - e).
\end{align}
Where the sigmoids used were defined as
\begin{equation}
	\sigma(x) = \frac{1}{e^{-\alpha x} + 1},
\end{equation}
with the $\alpha$ hyperparameter controlling the steepness of the function, recovering the exact form of the distribution for large values.
Therefore, we selected an $\alpha$ value of 50 in order to obtain a smooth enough $\tilde{p}_X$ that still presentes all the main fetures of the real distribution, as can be seen in Fig. \ref{fig:TrainRes}.
This allowed us to train an autoregressive neural spline model~\cite{durkan2019} to map a 2D diagonal gaussian into $\tilde{p}_X$ by minimising the standard reverse KL divergence of the model
\begin{equation}
	\mathcal{L}(\boldsymbol{\theta}) = \langle \ln\abs{\det J_{T_{\boldsymbol{\theta}}}} - \ln p_X\circ T_{\boldsymbol{\theta}} + S \circ T_{\boldsymbol{\theta}}\rangle_{p_Z},
\end{equation}
where $S(x)$ is a sum of softplus functions rising from the sigmoid contributions.
We obtained the final $T_{\boldsymbol{\theta}}(\vb{z}; \tau)$ by an unsupervised training where new samples were generated, at every step, from a set of $\tau$ uniformly distributed in the interesting domain of $G$ and used to estimate $\mathcal{L}$ as an average. Finally, the loss minimisation was carried out using standard stochastic gradient descent methods using a learning rate of $10^{-4}$, reduced to $10^{-5}$ after thirty thousand steps.

\section{Application and Discussion}

\begin{figure}[t!]
	\centering
	\includegraphics[width=\linewidth]{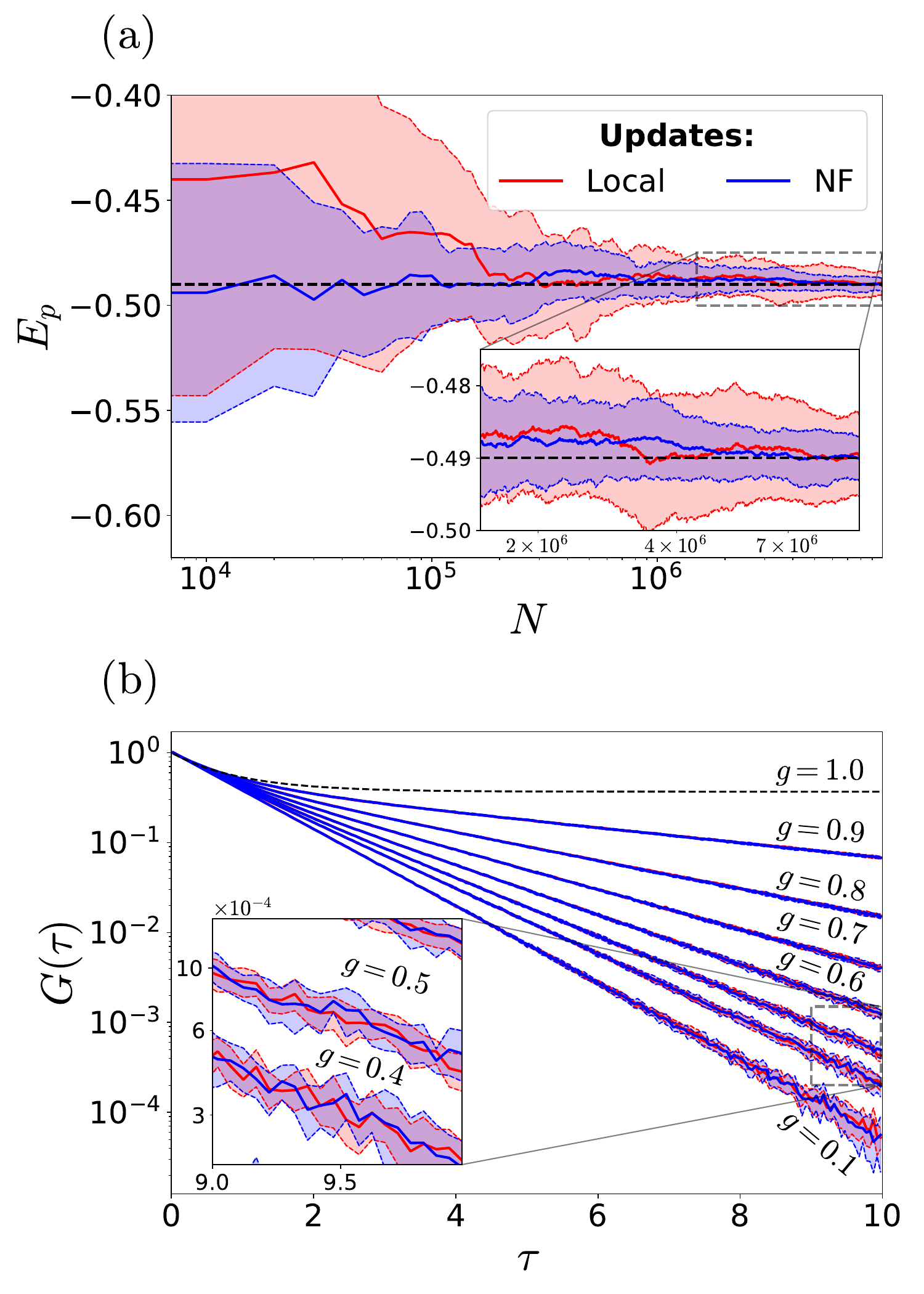}
	\caption{Comparison between standard and NF-augmented DMC for the estimate of $E_p$ and $G(\tau)$. (a) $E_p$ at $g=0.7$ as a function of the chain's steps; the bold line marks the mean value evolution, the coloured area represents the error, and the black dashed line indicates the exact value. (b) Green's function estimate (in logarithmic scale) as a function of $\tau$ for different coupling strength $g$. Lines and colours have the same meaning as in panel (a).}
	\label{fig:EgNeural}
\end{figure}

The developed NF-based architecture was integrated into the set of local updates in the Holstein model, establishing an NF-augmented Markov Chain. Subsequently, the statistical properties of this augmented chain were systematically compared against those of the standard one. Extensive data on observables and correlation times were gathered for various coupling strengths ($g$). Each run involved a set of 13 parallel chains, accumulating 10$^7$ steps.
The utilisation of separate chains facilitated the extraction of uncorrelated estimates, enabling the assessment of errors on individual observable values. This approach offered a direct means of comparing the performance of the two methodologies by analysing the evolution of mean values and errors throughout the chain.

\begin{figure}[t]
	\centering
	\includegraphics[width=\linewidth]{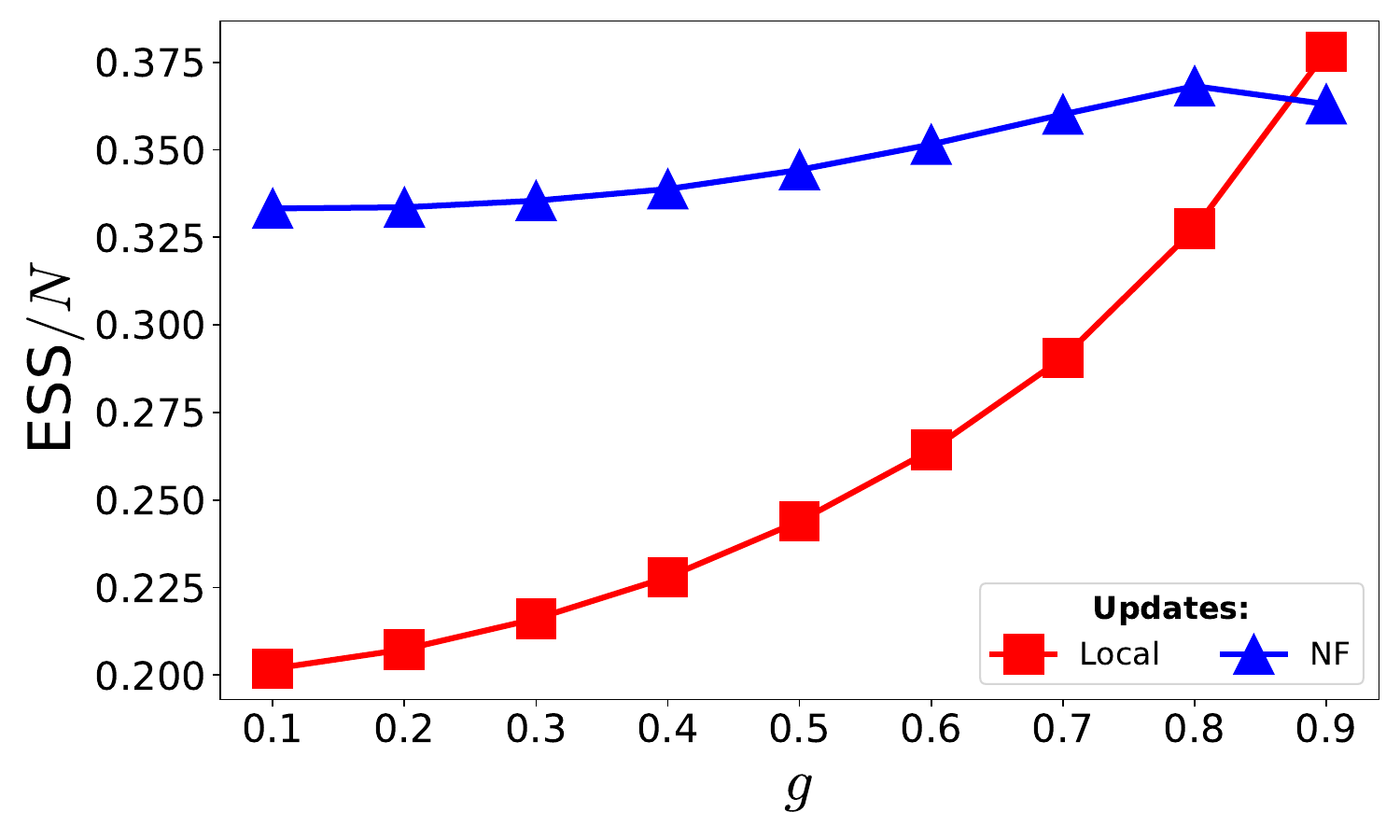}
	\caption{Effective sample size of the two different Markov Chains compared at different coupling constants. The algorithm using the NF global update remains superior along the whole line, showing a reduction in performances only upon reaching $g$ close to the model instability given by $G(\tau)$ becoming the identity.}
	\label{fig:ESS}
\end{figure}

\subsection{Statistical performances}

The evolution of the polaron energy ($E_p$) for a specific coupling strength ($g$), as depicted in Fig.\ref{fig:EgNeural}(a), clearly illustrates that the NF global update reduces the error in the estimate by approximately a factor of two compared to the local updates in standard DMC. Additionally, the Green's function presented in Fig.\ref{fig:EgNeural}(b) exhibits nearly identical performance for both types of updates across the entire interaction spectrum. This outcome is anticipated since the diagram's length $\tau$, serving as the correlated variable in $G(\tau)$, is taken as input from $p_W$ and is not sampled. Consequently, the correlation in the $\tau$ channel remains consistent with the standard case, with improvements primarily observed in phonon-related quantities.
To quantitatively evaluate the performance of the NF global updates, we present the effective sample size (ESS) as a function of the coupling strength $g$ in Fig.~\ref{fig:ESS}. The NF-based updates exhibit a noticeable reduction in the overall sample correlation, leading to an increase in ESS across all coupling regimes. This result underscores the effectiveness of the proposed model in accurately reconstructing the target density, thereby enhancing the overall performance of the NF-augmented Markov Chain.
However, a decrease in the performance of the NF model is observed in the strong coupling regime ($g = 0.9$). It is important to note that this anomalous behaviour is attributed to a peculiarity of the physical model rather than the NF architecture itself. Specifically, the single-site Holstein Green's function tends to a constant as $g$ approaches unity, leading to a flattening of the Green's function with increasing $g$ in Fig.~\ref{fig:EgNeural}(b). This causes every value of $\tau$ to have the same statistical weight, and very large values for $\tau$ start to appear in the chain, entering an extrapolation regime and lowering the model's performance. This is a system-specific issue that does not undermine the applicability of the NF model. It is always possible to adjust parameters, such as increasing $\varepsilon$ (currently set to 1), to artificially make high $\tau$ less probable without altering the observable estimate.

\subsection{Computational aspects}
The C++ package is constructed on top of the C++ API of PyTorch~\cite{pytorch}, while the plots were produced using Numpy~\cite{numpy} and Matplotlib~\cite{matplotlib}.
The training was carried on a single Nvidia GeForce GTX 1650 GPU and took a little less than an hour of computer time. The code is openly available in github~\cite{LLDMC}.
Additional details on the NF-DMC architecture is given in Appendix~\ref{sec:NfDes}.

As mentioned in the introduction, our code  was not optimised for performance since the focus was placed on exploring the possible statistical advantages of the methodology.
Therefore, no parallelisation or batching was implemented, and GPUs were only utilised for training, leading to longer sampling times at higher orders.
As a result, at present, the cost of our approach is approximately three times higher compared to the standard one at $g=0.1$ where an $\sim$ 40\% gain in effective sample size is observed, and become worse at larger $g$.

\section{Conclusions}
In conclusion, we presented a global sampling strategy for connected Feynman diagrams based on the use of Normalizing Flows, while integrating it into an NF-augmented Diagrammatic Monte Carlo framework.
Our devised procedure was tested on the Holstein polaron model, demonstrating superior statistical performance when compared to conventional DMC, albeit at an increased computational cost.
Diagrams are constructed using a bottom-up approach that sequentially samples interactions from a single unsupervised NF model trained solely utilising the weight function of the desired diagram type.
This strategy represents an initial attempt to enhance DMC using generative machine learning. It could potentially be extended to other model Hamiltonians by designing NF networks capable of representing the distinctive features of the targeted diagrams.
For instance, potential future developments include the incorporation of phonon momenta to account for dispersion and the utilisation of multiple NF models to describe systems with combined interactions~\cite{defilippis2021, Burovski2008}.
Moreover, its high flexibility, stemming from the ever-growing variety of NF architectures, the freedom to choose the order distribution, and the option to train without the need for a database, may attract interest from the broader quantum Monte Carlo community, facilitating its application to other perturbative methods.


\begin{acknowledgments}
	Support from the
	the Austrian Science Fund (FWF) project SFB-F81 project TACO is gratefully acknowledged.
	For open access
	purposes, the author has applied a CC BY public copyright license to
	any author accepted manuscript version arising from this submission.
	This work was partially supported by the NextGenerationEU-Piano Nazionale Resistenza e Resilienza (PNRR) CN-HPC grant no. (CUP) J33C22001170001, SPOKE 10.
	The computational results have been achieved using the Vienna Scientific Cluster (VSC). Fruitful discussion with Andrey Mishchenko and Stefano Ragni are greatly acknowledged.
\end{acknowledgments}


\appendix

\section{Description of the order distribution}
\label{App:DesOrDis}
\begin{figure}[t]
	\centering
	\includegraphics[width=\linewidth]{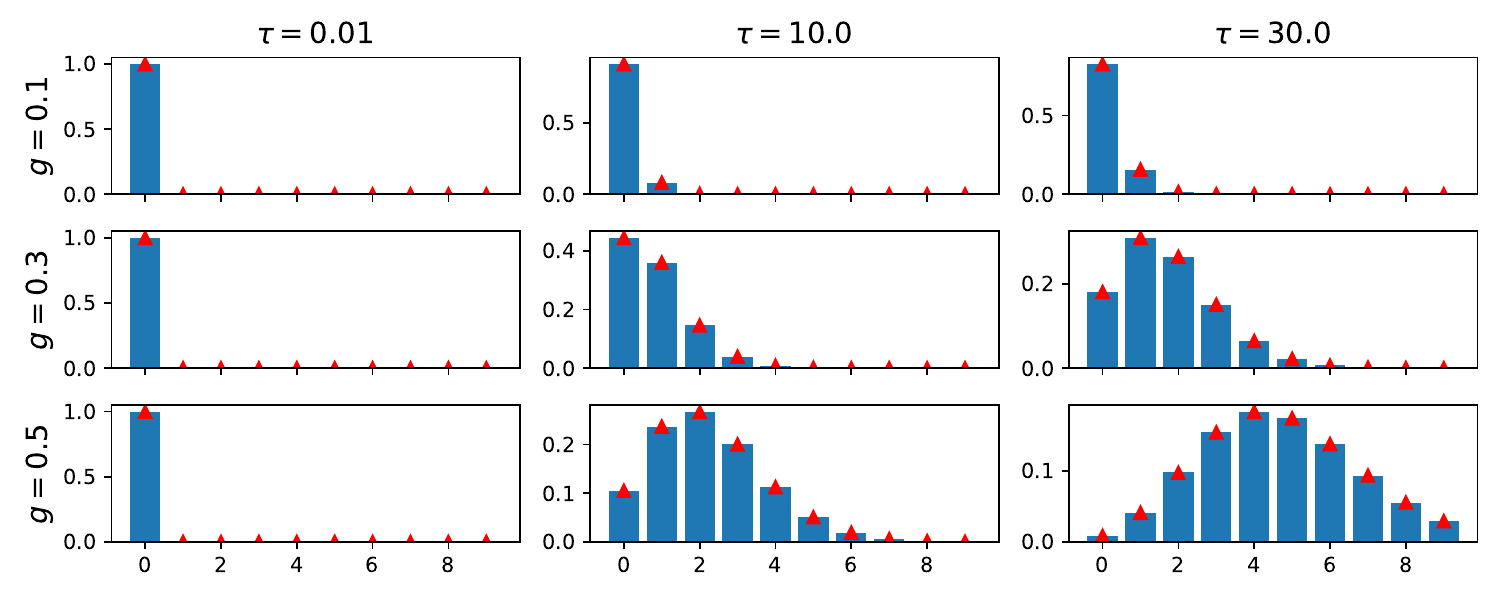}
	\caption{Test of the obtained $p_n(n\vert \tau, g)$ (red triangles) on the collected DMC distribution (blue histograms) at different $\tau$ and $g$ values. A perfect overlap of the analytic form used in this work and the raw data is clearly seen.}
	\label{fig:enter-label}
\end{figure}

The form of the distribution used to sample the order $p_n(n\vert \tau, g)$ was selected to best fit the order statistic collected from a set of DMC runs performed at different electron-phonon coupling strength $g$.
The data collected showed a form nearly identical to the one of a Poisson distribution for all the $\tau$ and $g$ values investigated, leading us to the following choice for the density used during the work
\begin{equation}
	p_n(n\vert \tau, g) = \frac{\lambda(\tau, g)^n}{n!}e^{-\lambda(\tau, g)}.
\end{equation}
To complete the distribution we noticed how the normalisation constant $e^{\lambda(\tau, g)}$ must coincide with the final result of the integration in Eq. \ref{eq:Green} reported here
\begin{equation}
	\begin{aligned}
		\label{eq:GreenA}
		G(\tau) & = e^{-\epsilon\tau}\sum_{n=0}^{\infty} \int_0^{\tau}\text{d}x_1\int_{x_1}^{\tau}\text{d}x_2\cdots\int_{x_{2n-1}}^{\tau}\text{d}x_{2n} W(\vb{x}) \\ &= e^{-\epsilon\tau - \lambda(\tau, g)} \sum_{n=0}^\infty p_n(n\vert \tau, g).
	\end{aligned}
\end{equation}
Therefore, by substituting the known result for $G(\tau)$~\cite{mahanManyParticlePhysics2000} and using the fact that $p_n$ is normalized one would obtain
\begin{equation}
	\exp[-\epsilon\tau - g^2(\tau - 1 + e^{-\tau})] = \exp[-\epsilon\tau - \lambda(\tau, g)],
\end{equation}
where $\Omega$ was set as unit of measure for the energy.
At the end a form of the function $\lambda(\tau, g)$ was obtained that respected perfectly the behaviour of the distribution
\begin{equation}
	\lambda(\tau, g) = g^2(\tau - 1 + e^{-\tau}),
\end{equation}
which avoided the need of interpolating the means found by fit of the collected statistics to get an approximated version of the exact distribution.

\section{Description of the Normalizing Flow architecture}
\label{sec:NfDes}

\begin{figure*}[t]
	\centering
	\includegraphics[width=\linewidth]{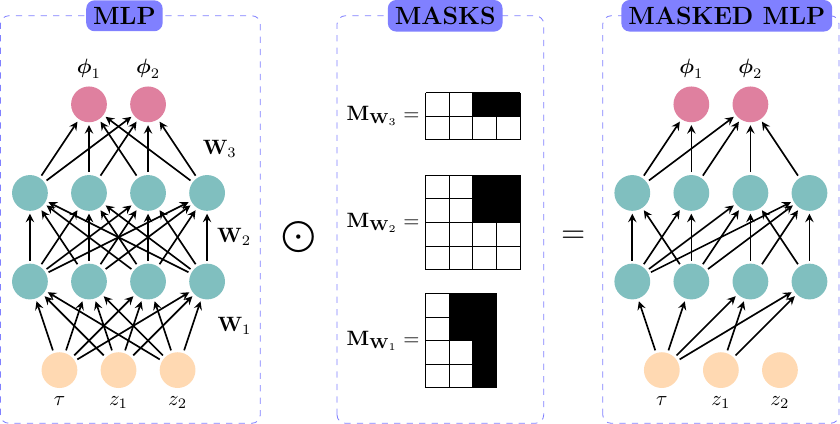}
	\caption{Graphical description of the masked MLP used in this work adapted from the idea in the MADE implementation~\cite{germain2015}. A fully connected network, on the left, gets truncated by multiplying the weight matrices $\mathbf{W}$ with masks $\mathbf{M}_{\mathbf{W}}$. The result is a simpler network, as can be seen on the right, where the j-th output, $\boldsymbol{\phi}_j$, clearly posses the wanted dependence on the input $\tau$ and $\mathbf{z}_{<j}$.}
	\label{fig:ModelSpec}
\end{figure*}

The model employed in this study is a conditional Normalizing Flow architecture using rational quadratic splines (RQS)~\cite{durkan2019} parametrized by the output of a masked multi-layer perceptron used as an autoregressive layer~\cite{papamakarios2017}.
To elucidate this architecture in greater detail, we present the mathematical formulation of $T_{\boldsymbol{\theta}}$ in an introductory way.

The parametric transformation constituting the flow is defined by its action on the components of the two-dimensional input vector $\mathbf{z}$, which is represented as follows:
\begin{equation}
	\label{eq:trans}
	\mathcal{R}_{\boldsymbol{\phi}_j}(z_j) = \begin{cases}
		g_{\boldsymbol{\phi}_j}(z_j) & z_j \in [-B, B]  \\
		z_j                          & \text{otherwise}
	\end{cases}.
\end{equation}
Where the function $g_{\boldsymbol{\phi}_j}$ is an invertible RQS parametrized by $\boldsymbol{\phi}_j$ containing the coordinates and derivatives, strictly positive, of each node.
Based on the number of nodes, $\mathcal{R}_{\boldsymbol{\phi}_j}$, is capable of representing more complex invertible transformations within the boundaries defined by $B$, which was fixed to 22 in order to operate within the Green's function's domain of interest $[0, 20]$.
Generally, the transformation is completed by employing specialized neural networks to compute the appropriate $\boldsymbol{\phi}_j$, based on the input $\mathbf{z}$ and our conditional parameter $\tau$, leading to the specific transformation mapping the initial distribution $p_Z$ to the target $\tilde{p}_X$.
Our model employed a multi-layer perceptron (MLP) to evaluate positions and derivative of 5 nodes per spline, leading to two $\boldsymbol{\phi}_j$ composed of 15 entries each for the 2D model used in the work.
In particular, a masked MLP (MMLP) was constructed based on the implementation of the Masked Autoencoder for Distribution Estimation (MADE)~\cite{germain2015}, showed in Fig.~\ref{fig:ModelSpec}, to compute $\boldsymbol{\phi}_j$ as a function of $\tau$ and $\mathbf{z}_{<j}$ as follows
\begin{align}
	\label{eq:cond}
	 & [\boldsymbol{\phi}_1(\tau), \boldsymbol{\phi}_2(\tau, z_1)] = \text{MMLP}(\mathbf{x}, \boldsymbol{\theta}), & \mathbf{x} = \text{concat}(\tau, \mathbf{z}).
\end{align}
Once Eq. \ref{eq:trans} and Eq. \ref{eq:cond} are coupled toghether they form a map acting on $\mathbb{R}^2$ parametrized by the network weights $\boldsymbol{\theta}$ and conditioned by $\tau$ that constitute the following conditional flow transform
\begin{align}
	 & R_{\boldsymbol{\theta}}(\mathbf{z}, \tau) = [\mathcal{R}_{\boldsymbol{\phi}_1(\tau)}(z_1), \mathcal{R}_{\boldsymbol{\phi}_2(\tau, z_1)}(z_2)],         \\
	 & \det J_{R_{\boldsymbol{\theta}}}(\mathbf{z}, \tau) = \mathcal{R}_{\boldsymbol{\phi}_1(\tau)}'(z_1) \mathcal{R}_{\boldsymbol{\phi}_2(\tau, z_1)}'(z_2).
\end{align}
Where we can see how the determinant of the Jacobian, $\det J_{R_{\boldsymbol{\theta}}}(\mathbf{z}, \tau)$, takes a simple form since the matrix is triangular.
The final transformation $T_{\boldsymbol{\theta}}(\bullet, \tau)$ presented as our model in Sec.~\ref{sec:FlowArch} was given by a composition of two $R_{\boldsymbol{\theta}}$, giving the following variational map
\begin{align}
	 & T_{\boldsymbol{\theta}}(\mathbf{z}, \tau) = R_{\boldsymbol{\theta}_1}(R_{\boldsymbol{\theta}_2}(\mathbf{z}, \tau), \tau),
	 & \boldsymbol{\theta} = [\boldsymbol{\theta}_1, \boldsymbol{\theta}_2].
\end{align}
The Jacobian determinant can then be evaluated using the chain rule of differentiation.

\bibliography{library.bib}

\end{document}